\documentclass[reprint,superscriptaddress,bibnotes,amsmath,amssymb,aps,pra]{revtex4-2}
\usepackage{graphicx}
\usepackage{dcolumn}	
\usepackage{bm}		
\usepackage{natbib}
\usepackage[colorlinks]{hyperref}
\usepackage[usenames,dvipsnames]{color}
\hypersetup{citecolor=Blue,linkcolor=Blue,urlcolor=Blue}

\begin{document}

\title{Frequency dependence of nonsequential double ionization\\of atoms in strong laser fields}

\author{Jan H. Thiede}
\affiliation{Fachbereich Physik, Philipps-Universit\"at Marburg, D-35032 Marburg, Germany}

\author{Jakub S. Prauzner-Bechcicki}
\email{jakub.prauzner-bechcicki@uj.edu.pl}
\affiliation{Jagiellonian University in Kraków, Faculty of Physics, Astronomy and Applied Computer Science,
Marian Smoluchowski Institute of Physics, Łojasiewicza 11, 30-348 Krakow, Poland}

\date{\today}

\begin{abstract}
The frequency dependence of (nonsequential) double ionization is studied in fully quantum mechanical calculations using a (1+1)-dimensional model atom. Other time-dependent effects such as the influence of the number of field cycles are also investigated. We present ionization yields and momentum distributions. The results are consistent with experimental and semi-classical data available in the literature, thus complementing them with a quantum mechanical description. 
\end{abstract}

\maketitle

\section{\label{sec:intro}Introduction}

Many intriguing and fascinating phenomena in nature thrive from correlations. One of the best examples from strong-field physics is so-called nonsequential double or multiple ionization (NSDI or NSMI) of atoms (see e.g. \cite{Faria11,becker2012theories} for a review). NSDI and NSMI processes are well described by a three-step model \cite{Corkum93} which is more commonly referred to as the rescattering mechanism. The whole story starts with electrons tunneling through the Stark barrier when the laser field is strong. Most of them will give rise to a single ionization signal. A small part of them, however, is accelerated in the laser field and driven back to the ion as the field reverses sign. Finally, in inelastic collisions, the returning electrons transfer some energy to the electrons remaining at the ion. As a result, the electrons form a transient, short-lived compound state which may decay through single, double or multiple ionization events or undergo a repetition of the rescattering cycle \cite{EckhardtSacha01,Camus12,QuantumModel08}. Additionally, the anti-correlated emission caused by a slingshot mechanism is to be expected~\cite{Katsoulis2018}.

In the following, we restrict ourselves to double ionization. There are two main double ionization channels \cite{QuantumModel08}. First, there are simultaneously escaping electrons. This channel is dominated by recollision-induced direct ionization (REDI) \cite{Weber00} events. Minor contributions to this channel come from the decays of autoionizing two-electron states that persisted through several field cycles and two-electron tunnel ionizations. Second, there are electrons following a consecutive path of escape, i.e. a path on which there is a time delay between the electrons' departures. In this channel, the independent escape of electrons, called sequential double ionization, is included. Another important contribution comes from indirect processes where the residual electron is excited to a higher bound state before leaving the atom (recollision-induced excitation plus subsequent field ionization (RESI) \cite{Feuerstein01}). One may also consider RESI-like processes in which the excited state persisted through several field cycles before ionization.

Experimentally, there are three main characteristics of NSDI. First, it was observed that in a certain laser intensity regime, double ionization yields are several orders of magnitude greater than expected from models based on a purely sequential ionization process. The observation is manifested by a pronounced knee structure in the yield vs. peak field amplitude plots \cite{lHuillier83a,Fittinghoff92,Walker94,Talebpour97,Larochelle98}, indicating production of ionized electrons in a coherent process at low intensities. Secondly, in that particular regime, the longitudinal momentum distributions of the doubly-charged ions show a double-hump structure \cite{Weber00}, suggesting that the escaping electrons leave the atom with similar non-zero longitudinal momenta. The depth of the minimum at zero momentum depends on the atomic species \cite{DeJesus04,Rudenko04}, implying non-negligible quantum mechanical effects. Finally, the highly-resolved two-electron momentum distributions show fingerlike or V-shaped structures, strengthening the observation that the two electrons typically leave the atom with similar nonzero momenta \cite{Staudte07,Rudenko07}. The latter features, i.e. the V-shaped structures, are sufficiently explained by REDI \cite{Faria11,ye2008}. However, the low momenta regions in the distributions, i.e. around $p_{1,\parallel}=p_{2,\parallel}=0$ (the minimum at zero ion momentum), are related to RESI \cite{Faria11,Bergues12}.

Many theoretical and experimental results concerning NSDI were published within the last few decades addressing different aspects of the phenomenon. 
On the experimental side, evidence for a frequency dependence of NSDI was already found early \cite{Kondo93,Larochelle98}. Later, the ratio of double to single ionization was studied for xenon over a broad range of frequencies and intensities \cite{Rudati04,Kaminski06,Gingras09}. Furthermore, the influence of the frequency on the longitudinal ion momentum distributions of argon and neon was investigated \cite{Herrwerth08,Alnaser08}. On the theoretical side, the frequency dependence of NSDI has been studied in several semiclassical calculations \cite{Larochelle98,Chen03,Emmanouilidou08,lu2019} and in a quantum mechanical one \cite{Corso00} using a mean-field model \cite{Watson96}. Recently, the frequency scaling of the NSDI yield was studied in joint experimental and semi-classical studies in Xe~\cite{wang2017} and Mg~\cite{kang2019}. Also the changes in the two-electron momentum distributions were addressed lately in joint experimental and theoretical studies, covering both the dependence on the ponderomotive energy $U_p=F_0^2/4\omega^2$~\cite{shaaran2019} or the pulse duration~\cite{Jiang2023}. 

In this paper, we present 
a detailed fully quantum mechanical \textit{ab initio} investigation of the frequency dependence of NSDI that includes both ionization yields \textit{and} momentum distributions. Other time-dependent effects such as the influence of the number of field cycles are also addressed. To achieve numerical feasibility, we use a (1+1)-dimensional model for helium (He) \cite{EckhardtSacha06,TimeResolved07,QuantumModel08,Suppression08,Eckhardt10,Efimov18}. We present ionization yields and momentum distributions for a broad range of frequencies and intensities. 
The report is organized as follows. In Section~\ref{sec:model}, the (1+1)-dimensional model atom is introduced. The numerics are briefly described in Section~\ref{sec:numerics}, followed by the presentation of the results in Section~\ref{sec:results}. Finally, in Section~\ref{sec:conclusion}, some conclusions are drawn.

\section{\label{sec:model}Model}

The exact theoretical description of NSDI relies on solving the full-dimensional Schr\"odinger equation \cite{Dundas99, Parker06, Armstrong11} or using $S$-matrix theory \cite{Becker05}. Since both methods are still computationally challenging, simplified models are often used instead. Confining the electrons to a common line parallel to the field polarization axis with coordinates $x$ and $y$ leads to the popular aligned-electron model \cite{Lappas98,Liu99}. However, such an approach suffers from the fact that the electron repulsion diverges on the diagonal $x=y$, thus preventing the symmetric electron escape with equal momenta observed in experiments \cite{Correlation00}. To remove this drawback, a reduction from six to three dimensions where the electronic center-of-mass motion is restricted to the field polarization axis was introduced and used in \cite{Ruiz06,Chen10}. Nevertheless, the computational cost of that method is still high.

Motivated by a classical analysis of the double ionization pathways \cite{EckhardtSacha01}, Eckhardt and Sacha introduced a (1+1)-dimensional model atom \cite{EckhardtSacha06} which proved to be capable of reproducing the aforementioned experimental key features of NSDI \cite{TimeResolved07,Suppression08,QuantumModel08,Eckhardt10,Efimov18}. A detailed discussion of the quantum mechanical implementation of the (1+1)-dimensional model is presented in \cite{TimeResolved07,QuantumModel08}, so we only point out the essential elements here. In Hartree atomic units and with the field interaction described in length gauge, the model's Hamiltonian reads
\begin{align}\label{eq:EShamiltonian}
H=\sum_{i=1}^2 \biggl(-\frac{1}{2}\frac{\partial^2}{\partial r_i^2}-\frac{2}{|r_i|}+\frac{\sqrt{3}}{2}F(t)r_i\biggr)\nonumber\\+\frac{1}{\sqrt{(r_1-r_2)^2+r_1r_2}}
\end{align}
where $r_1, r_2$ are the electron coordinates and $F(t)$ is the linearly polarized electric field of the laser. The factor $\sqrt{3}/2$ results from the model geometry. Replacing $F(t)r_i$ by $A(t)p_i$ in Eq.~(\ref{eq:EShamiltonian}) (where $A(t)=-\int_0^t F(t')\mathrm{d}t' $ is the vector potential), one obtains the Hamiltonian in velocity gauge which is more convenient for calculating momentum distributions \cite{QuantumModel08}.

The electric field of our laser pulse is given by \cite{Eckhardt10}
\begin{align}\label{eq:pulseshape1}
F(t)=\begin{cases}0,&t<0,\\F_0f(t),&0\leq t\leq T_p,\\0,&t>T_p\end{cases}
\end{align}
with ($t^*=t-T_p/2$)
\begin{align}\label{eq:pulseshape2}
f(t)=\cos\biggl(\frac{\pi t^*}{T_p}\biggr)\biggl[\cos\biggl(\frac{\pi t^*}{T_p}\biggr)\cos(\omega t^*+\varphi)\nonumber\\-\frac{1}{n_{c}}\sin\biggl(\frac{\pi t^*}{T_p}\biggr)\sin(\omega t^*+\varphi)\biggr]
\end{align}
where $F_0$, $\omega$, $\varphi$, $n_c$ and $T_p=2 \pi n_c/\omega$ are the peak amplitude, frequency, carrier-envelope phase, number of field cycles and pulse duration of the laser field, respectively. The rather complicated form of the pulse guarantees a vanishing pulse area for all values of $\varphi$ and hence a momentum distribution centered around the origin \cite{Rottke06}.

\section{\label{sec:numerics}Methods}

The numerical procedures for solving the Schr\"odinger equation of the (1+1)-dimensional model and for obtaining ionization yields and momentum distributions have been described elsewhere in detail \cite{QuantumModel08}, so we restrict ourselves to the basic ideas. First of all, the Coulomb terms in Eq.~(\ref{eq:EShamiltonian}) are smoothed with a cutoff parameter $\epsilon=0.6$~a.u., i.e. $1/|r|\rightarrow 1/\sqrt{r^2+\epsilon}$, leading to a ground state energy of the unperturbed atom of $E_g=-2.83$~a.u. The ground state is calculated by means of imaginary time evolution. The corresponding time-dependent Schr\"odinger equation is then solved on a square grid $[-L/2,L/2]\times [-L/2,L/2]$ with the split-operator method \cite{Feit82}. For a pulse with $n_c$ field cycles, the integration in time is carried out for $n_c+1$ cycles. The additional cycle is necessary to stabilize the system after the pulse is gone.

Using a grid of finite width physically corresponds to embedding the system in a box with infinitely high walls. As a consequence, the electron wave function is reflected when it reaches the edge of the grid. This problem is fixed by applying absorbing boundary conditions that become active at a certain distance $0\ll x_0<L/2$ from the center of the grid, i.e. the nucleus. We require $x_0$ and hence $L$ to be greater than the quiver radius $x_q=F_0/\omega^2$ of the classical motion to ensure that the main part of the wave function remains inside the grid.

For the yield calculation, the configuration space is partitioned into regions that correspond to the different ionization stages of the atom  and the probability fluxes across the boundaries are integrated in time \cite{Dundas99,TimeResolved07,QuantumModel08}. This procedure allows to identify two channels of double ionization: simultaneous double escape (SE) and consecutive double escape (CE)~\cite{QuantumModel08}. In the former case, the two-electron wave function directly passes from the neutral to the doubly-ionized atom regions whereas in the latter case, the singly-ionized atom regions are traversed. As pointed out in~\cite{QuantumModel08}, the quantitative predictions of the model are to some extent affected by the numerical values for positions of the borders between different regions. However, since we use the reduced-dimensionality model, the obtained yields may be treated as qualitative predictions only. Therefore, we adapt the same positions for borders as those used in previous studies~\cite{Dundas99,TimeResolved07,QuantumModel08,Suppression08,Eckhardt10,Efimov18}.

The calculation of momentum distributions is a little more complicated. For $|r_1|,|r_2|<x_0$ (inner region), the wave function is propagated with the full Hamiltonian (\ref{eq:EShamiltonian}) in velocity gauge. Beyond the absorbing boundaries (outer region), one electron ($|r_1|>x_0$, $|r_2|<x_0$ and $|r_1|<x_0$, $|r_2|>x_0$) or both electrons ($|r_1|,|r_2|>x_0$) are assumed to be ionized and simplified Hamiltonians without the respective Couloumb interactions are used for the propagation~\cite{Lein00}. Thus, the momentum information of the absorbed parts of the wave function is retained \cite{QuantumModel08}. The parts of the inner wave function that extend beyond an absorbing boundary are smoothly cut out and coherently added to the outer wave function in each time step~\cite{Lein00}. The final wave function of the system in momentum representation corresponds to the full momentum distribution of our system. Since we are interested in double ionization, we smoothly cut out the contributions of the neutral and singly-ionized atom from the final wave function, thus removing momenta close to the axes from the distribution. As in \cite{QuantumModel08}, only the parts of the wave function corresponding to $|r_1|,|r_2|>50$~a.u. are kept. To account for the discretized wave function and the finite experimental resolution, all momentum distributions are averaged with a Gaussian of width $\sigma_p=0.07$~a.u. Finally, we can obtain the longitudinal momentum distribution of the doubly-charged ion by defining the longitudinal ion momentum $p_{\parallel}^{2+}=-\sqrt{3}(p_1+p_2)/2$ and integrating the two-electron momentum distribution over the other new coordinate ($p_{\perp}^{2+}=-(p_1-p_2)/2$).

\section{\label{sec:results}Results}

Four frequencies will be considered in the following: $0.094$~a.u. (blue), $0.060$~a.u., $0.035$~a.u. and $0.023$~a.u. (all near-infrared), corresponding to wavelengths of 485 nm, 760 nm, 1313 nm and 2000 nm, respectively. Except for the second one, these are just the wavelengths studied in \cite{Alnaser08}. It is worth noting that the dipole approximation, i.e. assuming that the electric field is independent of position, is well justified in all cases since the wavelengths are much greater than the Bohr radius. The peak field amplitude $F_0$ is varied between $0.11$~a.u. and $0.57$~a.u., corresponding to intensities between $4.25\cdot 10^{14}$~W/cm$^2$ and $1.14\cdot 10^{16}$~W/cm$^2$. 

\subsection{Ionization yields}

\subsubsection{Frequency dependence}

\begin{figure}[tb]
\includegraphics{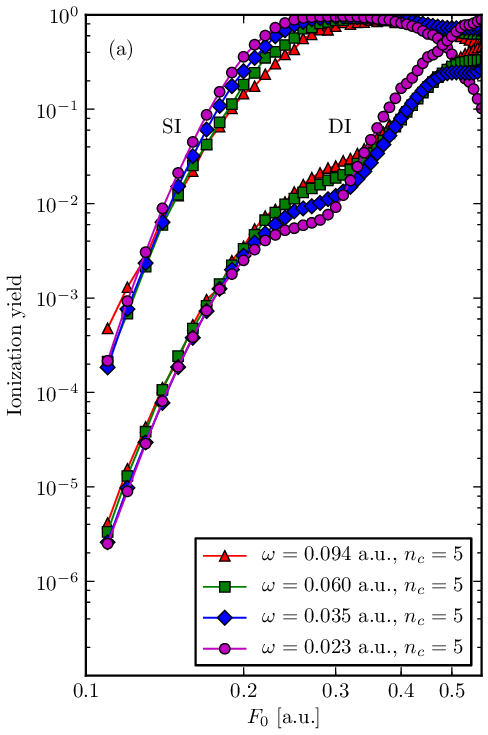}
\includegraphics{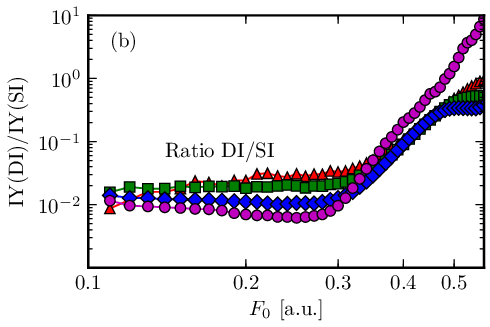}
\caption{\label{fig:IonYields}(color online) (a) Yields of single ionization (SI) and double ionization (DI) as a function of the peak field amplitude $F_0$ for $n_c=5$, $\varphi=0$ and different frequencies $\omega$. (b) Ratio of double to single ionization for the same parameters.}
\end{figure}

The single and double ionization yields (SI and DI, respectively) obtained for $n_c=5$, $\varphi=0$ and the four frequencies mentioned above are shown in Fig.~\ref{fig:IonYields}(a) as a function of the peak field amplitude. We first discuss them for a fixed frequency.

The numerical yields resemble the experimental ones \cite{Fittinghoff92,Kondo93,Walker94,Talebpour97,Larochelle98} in many ways. The knee structure in the DI yield as well as the saturation of the SI yield resulting from the depletion of neutral atoms in the interaction volume are clearly visible. The saturation field amplitude $F_{sat}$ is usually defined as the peak field amplitude for which the total ionization probability equals unity, i.e. \cite{Walsh93} 
\begin{align}\label{eq:deffsat}
\int_0^{T_p} W(F(t))\biggl|_{F_0=F_{sat}}\mathrm{d}t=1
\end{align}
where $W$ is the total ionization rate. Using the fact that the fraction of ionized atoms at the end of the pulse can be written as $P_{ion}(F_0) = 1 - \exp( - \int_0^{T_p} W(F(t))\mathrm{d}t)$ \cite{Scrinzi99}, we get $P_{ion}(F_{sat}) = 1 - 1/e \approx 0.632$. Hence, we can determine $F_{sat}$ approximately by inspecting the total ionization yield. 

In experiments, the interaction volume expands when the field amplitude is increased beyond $F_{sat}$, enabling ionization of more neutral atoms \cite{Mainfray91}. This effect obscures the saturation of SI above $F_{sat}$ and the depletion of singly-charged ions due to sequential DI for even higher field amplitudes. Instead, the SI yield increases as $F_0^{3}$ there \cite{Mainfray91,lHuillier83a,Walker94}, as expected for a Gaussian spatial intensity distribution. Here, we use the dipole approximation where the laser field is independent of position. Therefore, the considered interaction volume, i.e. the integration region, remains constant for all field amplitudes and as a consequence, the depletion of singly-charged ions becomes so significant above $F_{sat}$ that the increase of the SI yield changes into a decrease, producing a maximum at a certain field amplitude $F_{max}>F_{sat}$. $F_{max}$ coincides relatively well with the end of the knee, marking the onset of the regime where sequential DI dominates and doubly-charged ions are primarily formed through single ionization of singly-charged ions. 

Overall, the frequency dependence of the yields in Fig.~\ref{fig:IonYields}(a) is weak, as opposed to the strong dependence on the peak field amplitude. The yields are of the same order of magnitude for each of the ionic species and over the whole range of field amplitudes. However, the order with respect to $\omega$ differs. On the one hand, the SI yield increases with increasing $\omega$ for the lowest field amplitudes but this order reverses above $F_0\approx0.15$~a.u. On the other hand, the DI yield increases with increasing $\omega$ for nearly all values of $F_0$, with $\omega=0.023$~a.u. being the obvious exception to this trend (at higher field amplitudes, above $F_0\approx 0.3$~a.u.). The obtained results are in very good agreement with the data available in the literature~\cite{wang2017,kang2019}.

\begin{table}[b]
\caption{\label{tab:fmax}Saturation field amplitudes $F_{sat}$, field amplitudes $F_{max}$ for which the single ionization yield reaches its maximum and their difference $\Delta F=F_{max}-F_{sat}$ for $n_c=5$, $\varphi=0$ and different frequencies $\omega$.}
\begin{ruledtabular}
\begin{tabular}{cccc}
\textrm{$\omega$ [a.u.]}&
\textrm{$F_{sat}(\omega)$ [a.u.]}&
\textrm{$F_{max}(\omega)$ [a.u.]}&
\textrm{$\Delta F$ [a.u.]}\\
\colrule
0.094&0.27&0.35&0.08\\
0.060&0.25&0.34&0.09\\
0.035&0.23&0.31&0.08\\
0.023&0.22&0.31&0.09\\
\end{tabular}
\end{ruledtabular}
\end{table}

The values of $F_{sat}$ and $F_{max}$ are given in Table~\ref{tab:fmax}. As $\omega$ grows, $F_{sat}$ and $F_{max}$ move towards larger field amplitudes with their difference $\Delta F = F_{max}-F_{sat}$ remaining more or less constant. Consequently, the knee regime is shifted to higher field amplitudes with increasing $\omega$. We will track down the cause for this shift by investigating the influence of the pulse duration on the ionization yields in the next subsection. Here it is worth noting that the shift of the knee regime is responsible for the impression that the double ionization yield for $\omega=0.023$~a.u. rises much more steeply at the end of the knee than it does for the other frequencies.

The ratio of double to single ionization is shown in Fig.~\ref{fig:IonYields}(b) and is in agreement with experimental data~\cite{wang2017,kang2019}. In the knee regime, the dependence on the field amplitude is weak, giving rise to a plateau-like structure. For $\omega=0.060$~a.u., the ratio is practically constant, an observation also made in \cite{Walker94,Larochelle98}. 

\begin{figure}[tb]
\includegraphics{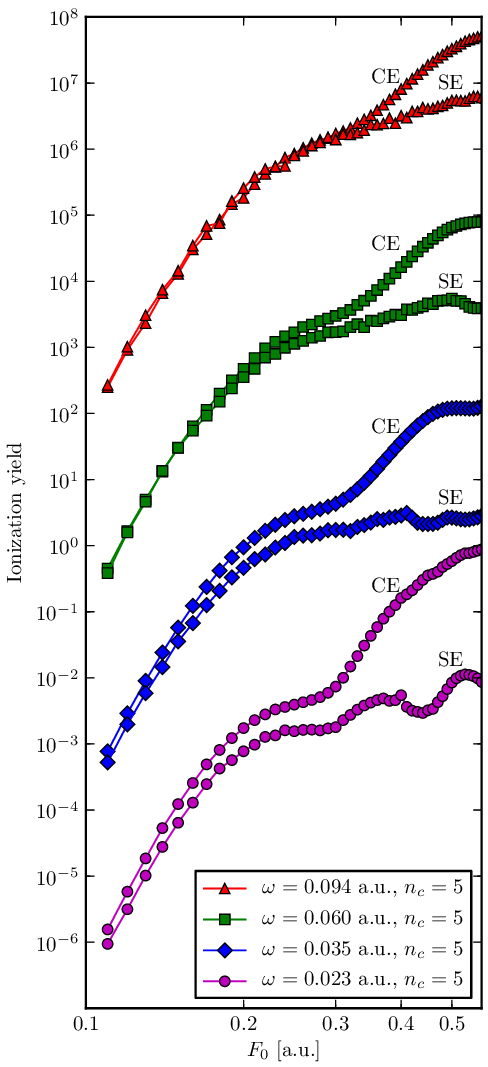}
\caption{\label{fig:IonYieldsCDIvsSDI}(color online) Yields of consecutive double ionization (CE) and simultaneous double ionization (SE) as a function of the peak field amplitude $F_0$ for $n_c=5$, $\varphi=0$ and different frequencies $\omega$. The yields are artificially separated by a factor of 500.}
\end{figure}

In Fig.~\ref{fig:IonYieldsCDIvsSDI}, we compare the CE and SE yields. For the sake of clarity, the yields belonging to different $\omega$ are artificially separated by a factor of 500. Nevertheless, we observe that they all are of the same order of magnitude below saturation, with the difference becoming more pronounced as $\omega$ grows. For the three lowest frequencies, CE slightly dominates over SE below saturation. This is mainly a consequence of the size and shape of the cruciform region in the configuration space that represents the neutral atom \cite{QuantumModel08}. By adjusting its borders, we can directly influence the magnitude of each channel, as shortly discussed in Section~\ref{sec:numerics}. The SE yields increase by about three orders of magnitude between 0.11 a.u. and 0.20 a.u. and eventually saturate. In the saturation regime, the slope of the yield (and hence the yield itself) decreases with decreasing $\omega$. The dependence of the CE yields on $\omega$ is even weaker. However, in contrast to the SE case, the CE yields display a knee structure since they contain both RESI and sequential DI, and the latter dominates the DI yields above $F_{max}$. It should be noted that there are still enough neutral atoms to allow for NSDI, so the simultaneous yields continue to grow above the knee. Below saturation, the contributions of sequential DI are negligible and the yield in the consecutive channel results mainly from RESI-like processes.

\subsubsection{Influence of the pulse duration}\label{sec:influence}

\begin{figure}[tb]
\includegraphics{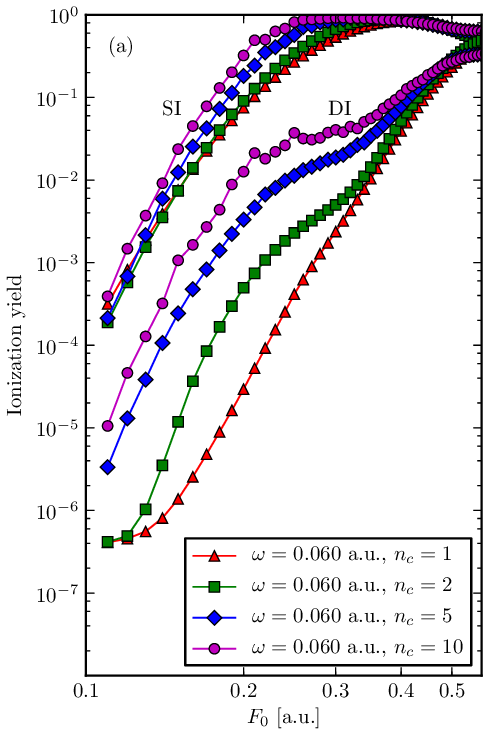}
\includegraphics{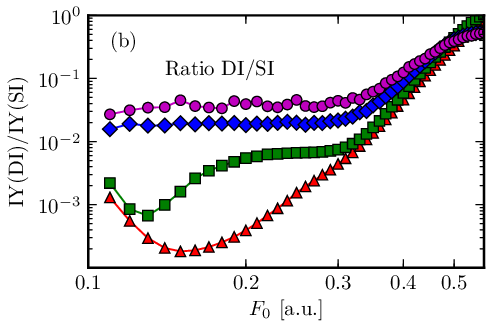}
\caption{\label{fig:IonYields_diffnc}(color online) (a) Yields of single ionization (SI) and double ionization (DI) as a function of the peak field amplitude $F_0$ for $\omega=0.060$~a.u., $\varphi=0$ and different numbers of field cycles $n_c$. (b) Ratio of double to single ionization for the same parameters.}
\end{figure}

In all yield calculations up to now, a laser pulse with five field cycles was used. Fixing the number of field cycles $n_c$ and changing the frequencies results in different pulse durations $T_p=2\pi n_c/\omega$, i.e. the values of $T_p$ for the pulses used in the previous subsection ranged from $8$~fs ($\omega=0.094$~a.u.) to $34$~fs ($\omega=0.023$~a.u.). It was argued in \cite{Larochelle98} that the pulse duration has a negligible influence on the NSDI yields since the only significant time scale is the ejection of the later-rescattering electron which is on the order of a field cycle or less. While this reasoning might be true for long pulses ($n_c\gg 1$), it clearly fails for few-cycle pulses. For these, the pulse duration is expected to have a major impact on the yields and to determine the dominating channel of double ionization. To infer the $T_p$ influence on NSDI, we consider three cases for the calculations: (i) varying $T_p$ by keeping $n_c$ fixed and varying $\omega$, (ii) varying $T_p$ by keeping $\omega$ fixed and varying $n_c$, (iii) fixing $T_p$ by using values of $n_c$ and $\omega$ so that the ratio $n_c/\omega$ is constant. Case (i) was already investigated in the previous subsection, so we now turn to the remaining ones. 

\begin{figure}[tb]
\includegraphics{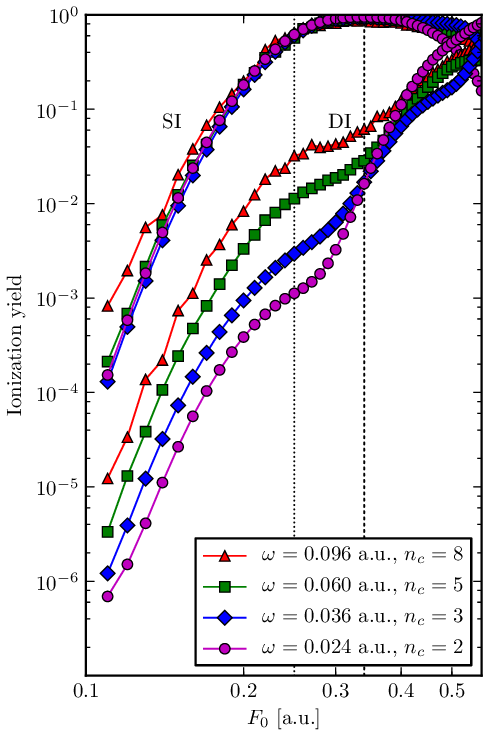}
\caption{\label{fig:IonYields_diffnc_diffomega}(color online) Yields of single ionization (SI) and double ionization (DI) as a function of the peak field amplitude $F_0$ for $\varphi=0$ and a constant pulse duration of $T_p=13$~fs. All four single ionization yields saturate at $F_{sat}=0.25$~a.u. (dotted line) and exhibit a maximum at $F_{max}=0.34$~a.u. (dashed line).}
\end{figure}

\begin{table}[b]
\caption{\label{tab:fmax2}Saturation field amplitudes $F_{sat}$, field amplitudes $F_{max}$ for which the single ionization yield reaches its maximum and their difference $\Delta F=F_{max}-F_{sat}$ for $\omega=0.060$~a.u., $\varphi =0$ and different numbers of field cycles $n_c$.}
\begin{ruledtabular}
\begin{tabular}{cccc}
\textrm{$n_c$}&
\textrm{$F_{sat}(\omega)$ [a.u.]}&
\textrm{$F_{max}(\omega)$ [a.u.]}&
\textrm{$\Delta F$ [a.u.]}\\
\colrule
$1$&0.32&0.41&0.09\\
$2$&0.29&0.38&0.09\\
$5$&0.25&0.34&0.09\\
$10$&0.23&0.31&0.08\\
\end{tabular}
\end{ruledtabular}
\end{table}

For case (ii), we choose $\omega=0.060$~a.u., $\varphi=0$ and $n_c\in\lbrace1,2,5,10\rbrace$, and the relevant ionization yields are shown in Fig.~\ref{fig:IonYields_diffnc}(a). Obviously, NSDI is enhanced in a pulse with more field cycles since the height of the knee increases. Such a trend can be easily explained with the three-step model. Since the number of field extrema grows with the number of field cycles, the number of ionized electron wave packets which can take part in rescattering events leading to NSDI increases as well. These wave packets can also move and interfere with each other for a longer time with increasing $n_c$ which manifests in the distortion of the yields. This distortion was also visible in the time-dependent probability fluxes studied in \cite{TimeResolved07} where an eight-cycle pulse with a trapezoidal envelope was used. For better illustration of the NSDI enhancement, the ratio of double to single ionization is shown in Fig.~\ref{fig:IonYields_diffnc}(b). We find that it increases with increasing pulse duration, at least before sequential double ionization starts to dominate. Moreover, the plateau region grows wider. These trends are in agreement with the experiment in \cite{Bhardwaj01} where the ratio was measured for neon and a Ti:sapphire laser ($\omega=0.057$~a.u.) with pulse durations of $12$~fs and $50$~fs.

The $F_{sat}$ and $F_{max}$ obtained in the discussed instance are given in Table~\ref{tab:fmax2}. Clearly, they both decrease with growing $n_c$. This trend was also found in \cite{Corso00} and can be explained as follows. As apparent from Eq.~\ref{eq:deffsat}, the saturation field amplitude $F_{sat}$ depends on the frequency and the number of field cycles. In the multiphoton case \cite{Mainfray91}, the static-field ionization rate $W(F_0)$ increases monotonically with $F_0$. Provided that $F_0$ is not too big, this also applies to the tunneling case \cite{ADK86}. Let $F(t)=F_0\Theta (T_p - t)$. Then Eq.~\ref{eq:deffsat} reduces to $W(F_{sat})T_p=1$ and $F_{sat}$ decreases if the pulse duration is increased. Now let $F(t)=F_0 \sin(\omega t)\Theta (T_p -t)$. In this case, $\omega$ and $n_c$ can be scaled out of the integral and Eq.~\ref{eq:deffsat} can be written as $ (T_p/2\pi) \int_0^{2\pi} W(F_{sat} \sin(\tau))\mathrm{d}\tau=1$. Again, $F_{sat}$ decreases with increasing $T_p$. Finally, let $F(t)$ be defined by Eq.~\ref{eq:pulseshape1}. Taking $x=(t-T_p/2)/T_p$ leads to $T_p\int_{-1/2}^{1/2}W(F_{sat}f(x))\mathrm{d}x =1$. Hence, $F_{sat}$ decreases with increasing $T_p$ -- the longer the pulse, the smaller the field amplitude where the ionization probability reaches unity and the knee regime begins.

\begin{table}[tb]
\caption{\label{tab:fmax3}Saturation field amplitudes $F_{sat}$, field amplitudes $F_{max}$ for which the single ionization yield reaches its maximum and their difference $\Delta F=F_{max}-F_{sat}$ for $\varphi =0$ and $n_c/\omega=1/0.012=\mathrm{const.}$}
\begin{ruledtabular}
\begin{tabular}{cccc}
\textrm{($n_c$, $\omega$ [a.u.])}&
\textrm{$F_{sat}(\omega)$ [a.u.]}&
\textrm{$F_{max}(\omega)$ [a.u.]}&
\textrm{$\Delta F$ [a.u.]}\\
\colrule
(8, 0.096)&0.25&0.34&0.09\\
(5, 0.060)&0.25&0.34&0.09\\
(3, 0.036)&0.25&0.34&0.09\\
(2, 0.024)&0.25&0.34&0.09\\
\end{tabular}
\end{ruledtabular}
\end{table}

The above argument can also explain the shift of the knee regime for constant $n_c$ and decreasing $\omega$ which was visible in Fig.~\ref{fig:IonYields}(a), i.e. case (i). If the dependence on the pulse duration was the sole reason for the shift, $F_{sat}$ and $F_{max}$ should be constant for constant $n_c/\omega$. To investigate this, we took ($n_c=5$, $\omega=0.060$~a.u.) as a reference and calculated the yields for ($n_c=8$, $\omega=0.096$ a.u.), ($n_c=3$, $0.036$~a.u.) and ($n_c=2$, $\omega=0.024$~a.u.). Note that the frequencies were slightly adjusted in order to be able to use an integer number of field cycles. The respective single and double ionization yields are compared in Fig.~\ref{fig:IonYields_diffnc_diffomega}. In all four cases, we find $F_{sat}=0.25$~a.u. and $F_{max}=0.34$~a.u. This result confirms that the dominance of the DI yield for $\omega=0.023$~a.u. over the DI yields for the other frequencies at the end of the knee region observed in Fig.~\ref{fig:IonYields}(a) stems from the shift of $F_{sat}$ caused by the change in the pulse duration. Since we now have two quantities acting in the same direction, i.e. a large (small) frequency combined with a large (small) number of field cycles, the differences between the DI yields are more pronounced than in Fig.~\ref{fig:IonYields}(a). 

\subsubsection{Rate equations}

\begin{figure}[tb]
\centering\includegraphics{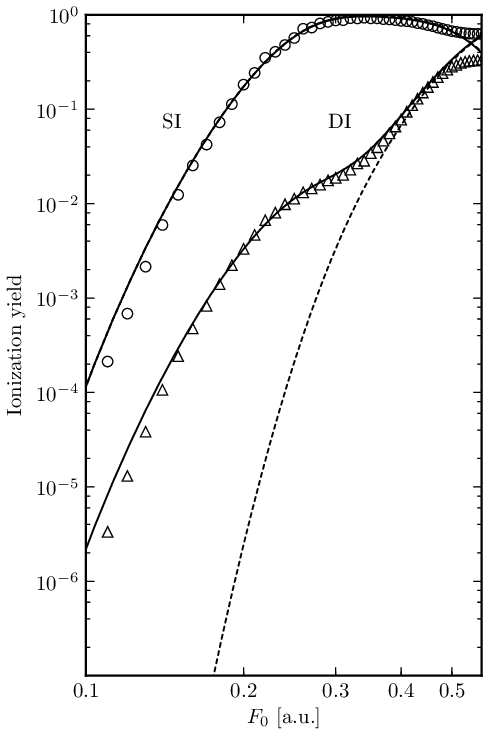}
\caption{\label{fig:yields_ADK} Comparison of the numerical ionization yields to the ones obtained by integrating the rate equations (\ref{eq:rates}) for $n_c=5$, $\omega = 0.060$~a.u. and $\varphi=0$. The sequential ionization rates ($W_{01}$ and $W_{12}$) are approximated by the ADK formula, the nonsequential one is approximated by $W_{02}=0.019\cdot W_{01}$ (solid lines) or set to zero (dashed lines).}
\end{figure}

As it was done in the first report on NSDI \cite{lHuillier83a}, we can consider rate equations to understand the ionization yields quantitatively. Assuming that the total number of particles is conserved, these equations are
\begin{subequations}\label{eq:rates}
\begin{align} 
\frac{\text{d}P_0}{\text{d}t} &= -W_{01}P_0-W_{02}P_0,\\
\frac{\text{d}P_1}{\text{d}t} &= \phantom{-}W_{01}P_0-W_{12}P_1,\\
\frac{\text{d}P_2}{\text{d}t} &= \phantom{-}W_{02}P_0+W_{12}P_1,
\end{align}
\end{subequations}
where $P_i(t)=N_i(t)/N_0(t=0)$ is the fraction of He$^{i+}$ at time $t$ and $W_{ij}$ is the rate of ionization from He$^{i+}$ to He$^{j+}$ ($i=0,1;j=1,2$). Reasonable initial conditions are $P_0(t=0)=1$ and $P_1(t=0)=P_2(t=0)=0$, i.e. there are only neutral atoms before the laser hits the interaction volume.

\begin{figure*}[tb]
\centering\includegraphics{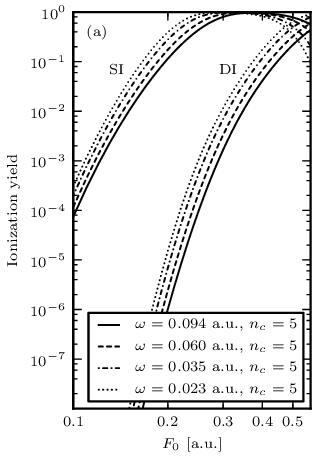}\includegraphics{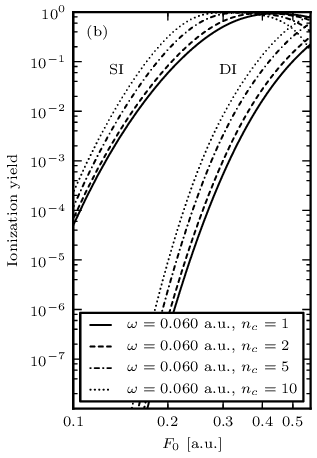}\includegraphics{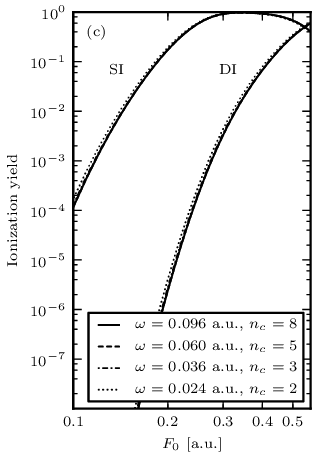}
\caption{\label{fig:yields_ADK_onlysequential}Solutions of the rate equations (\ref{eq:rates}) for different frequencies $\omega$ and numbers of field cycles $n_c$ as a function of the peak field amplitude. The sequential ionization rates ($W_{01}$ and $W_{12}$) are approximated with the ADK formula, the nonsequential one ($W_{02}$) is set to zero.}
\end{figure*}

In the tunneling regime, the sequential ionization rates can be approximated by the formula of Ammosov, Delone and Krainov (ADK) \cite{ADK86} which, for $s$-like states, only depends on the instantaneous field amplitude $F(t)$ and an effective quantum number $n^*=Z/\sqrt{2E_I}$ containing the charge of the residual ion $Z$ and the ionization energy $E_I$. Although the ADK formula is only valid for $\gamma\ll 1$ \footnote{The Keldysh parameter $\gamma=\sqrt{E_I/2U_p}=\sqrt{2E_I}\omega/F_0$ \cite{Keldysh65} relates the minimum energy needed for ionization, i.e. the ionization energy $E_I$, to the maximum energy a free electron can acquire in the field, i.e. $2U_p$ (where $U_p=F_0^2/4\omega^2$ is the ponderomotive energy).} and $n^*\gg 1$, it is successfully used to fit the sequential parts of experimental ionization yields even if these requirements are not fulfilled (see e.g. \cite{Fittinghoff92,Walker94,Kondo93,Larochelle98}). Note that we calculate the ionization yields at a finite distance from the nucleus, so ionization is, compared to an infinite grid, easier. To account for this fact, the ionization energies \footnote{One obtains the ionization energies of the model atom by calculating the field-free ground states of the Hamiltonian (\ref{eq:EShamiltonian}) and the Hamiltonian of the singly-ionized atom (let $r_2\rightarrow\infty$ in (\ref{eq:EShamiltonian})). For $\epsilon=0.6$~a.u., the ionization energies are $E_I=0.98$~a.u. and $E_I^+=1.85$~a.u.} entering the ADK formula are slightly lowered in the following, i.e. the effective quantum numbers $n_{01}^*$ (He $\rightarrow$ He$^+$) and $n_{12}^*$ (He$^{+}$ $\rightarrow$ He$^{2+}$) are slightly increased.

A general expression for the nonsequential rate $W_{02}$ is not known. However, we can exploit the fact that the ratio $r$ of double to single ionization is almost constant in the knee regime by setting $W_{02}=r\cdot W_{01}$. For a five-cycle pulse with $\omega=0.060$~a.u. and $\varphi=0$, we find $r\approx0.019$ from Fig.~\ref{fig:IonYields}(b). In Fig.~\ref{fig:yields_ADK}, the numerical ionization yields for these parameters are compared to the solution of the rate equations (\ref{eq:rates}). No attempt was made to optimize $n_{01}^*$ and $n_{12}^*$, e.g. in the least-square sense. Nevertheless, the agreement is good. In contrast to this, the result for $W_{02}=0$ (dashed lines) strongly underestimates the double ionization yield at low field amplitudes. 

Some of the observations from the previous subsections can already be understood without the nonsequential contribution, i.e. for  $W_{02}=0$. Figs.~\ref{fig:yields_ADK_onlysequential}(a)--(c) show the purely sequential SI and DI yields obtained by solving the rate equations (\ref{eq:rates}) for the cases (i)--(iii) (see sec.~\ref{sec:influence}), respectively. The SI maximum exhibits the same behavior as in Figs.~\ref{fig:IonYields}(a), \ref{fig:IonYields_diffnc}(a) and \ref{fig:IonYields_diffnc_diffomega}, i.e. it only depends on the pulse duration. As $F_{max}$ marks the regime where sequential DI starts to dominate the DI channel, the DI yields obtained from the rate equations should resemble its behavior. Indeed, the changes in the DI yields with frequency and number of cycles, Figs.~\ref{fig:yields_ADK_onlysequential}(a) and (b), respectively, agree very well with the previously discussed shifts in position of $F_{max}$. This observation is pinpointed by the unchanged DI yields for constant $\omega/n_c$, see Fig.~\ref{fig:yields_ADK_onlysequential}(c).

\subsection{Momentum distributions}

\begin{figure}[tb]
\centering\includegraphics{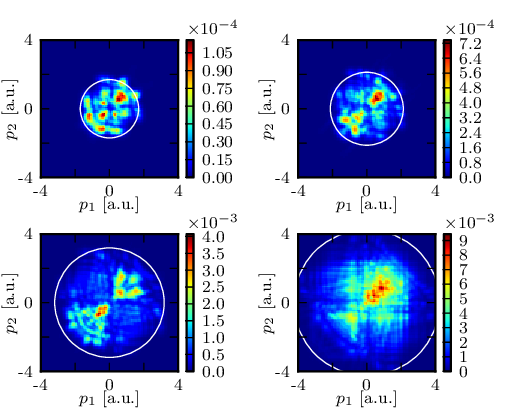}
\caption{\label{fig:ElMomDis1} (color online) Two-electron momentum distributions for $n_c=5$, $\omega=0.094$~a.u., $\varphi=0$ and the field amplitudes (from upper left to lower right) $F_0=0.16$~a.u., $F_0=0.20$~a.u., $F_0=0.30$~a.u. and $F_0=0.40$~a.u. The white circles correspond to the classical kinematical constraint $\pm2\sqrt{U_p}$ for the electrons' momenta.}
\end{figure}

\begin{figure}[tb]
\centering\includegraphics{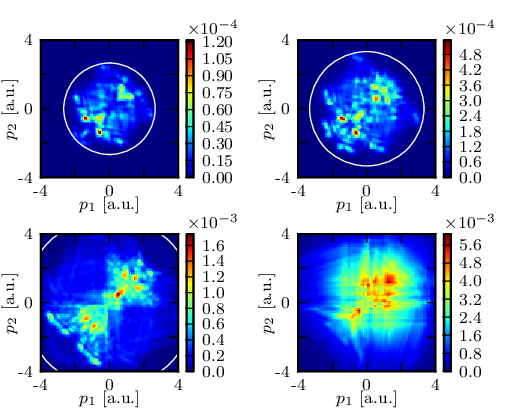}
\caption{\label{fig:ElMomDis2} (color online) 
Same as Fig.~\ref{fig:ElMomDis1}, but for $\omega = 0.060$~a.u.
}
\end{figure}

\begin{figure}[tb]
\centering\includegraphics{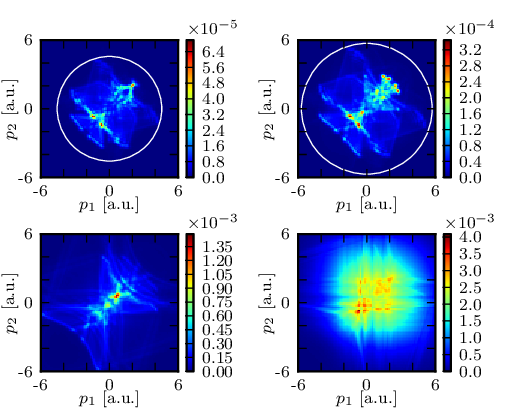}
\caption{\label{fig:ElMomDis3} (color online) 
Same as Fig.~\ref{fig:ElMomDis1}, but for $\omega = 0.035$~a.u.
}
\end{figure}

\begin{figure}[tb]
\centering\includegraphics{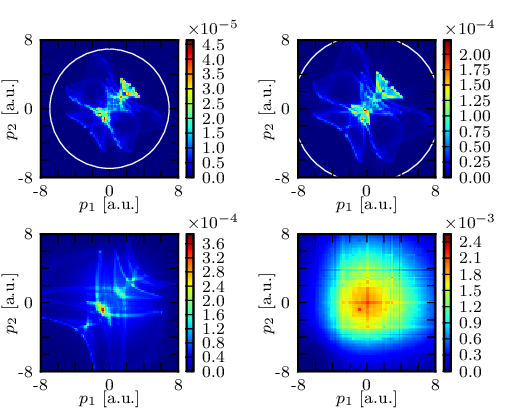}
\caption{\label{fig:ElMomDis4} (color online) 
Same as Fig.~\ref{fig:ElMomDis1}, but for $\omega = 0.023$~a.u.
}
\end{figure}

To make our study comprehensive, we discuss the two-electron momentum distributions,
shown in Figs.~\ref{fig:ElMomDis1}--\ref{fig:ElMomDis4}. For each frequency, we picked four field amplitudes to address the different regimes of double ionization visible in Fig.~\ref{fig:IonYields}(a): $F_0=0.16$~a.u. (nonsequential regime, below the knee), $F_0=0.20$~a.u., $F_0=0.30$~a.u. (saturation regime, the knee) and $F_0=0.40$~a.u. (sequential regime, above the knee).

As can be inferred from Figs.~\ref{fig:ElMomDis1}--\ref{fig:ElMomDis4}, the width of the distributions grows with increasing $F_0$ and decreasing $\omega$. This observation is related to the increase of the ponderomotive energy $U_p=F_0^2/4\omega^2$ and in line with the experiments (see e.g. \cite{Weber00,Alnaser08,Herrwerth08,shaaran2019}). For comparison, the parallel momentum threshold of $\pm 2 \sqrt{U_p}$ predicted by the rescattering mechanism is also included in the figures. It corresponds to a free electron leaving the ion parallel to the field polarization axis with the maximum energy it can classically acquire in the field ($2U_p$). Due to the choice of the laser pulse shape, we obtain zero drift momentum and the boundary set by the threshold is just a circle centered around the origin \cite{Feuerstein01}. In all cases, the two-electron momentum distributions lie well within these circles.

The nonsequential and the sequential regimes of double ionization can be rather easily identified just by inspection of the distributions. In the nonsequential regime, we observe the typical fingerlike or V-shaped structures indicative of correlated simultaneous double ionization \cite{TimeResolved07,QuantumModel08} which were also found experimentally \cite{Staudte07,Rudenko07}. In the sequential regime, the correlation of electron momenta is lost \cite{Correlation00} and all quadrants in the two-electron momentum distributions are populated.

\begin{figure}[tb]
\centering\includegraphics{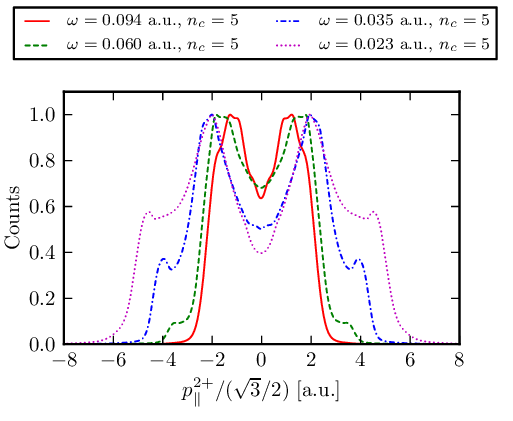}
\caption{\label{fig:IonMomDis_averaged} (color online) Longitudinal ion momentum distributions for $n_c=5$, $F_0=0.16$~a.u. and different frequencies $\omega$, averaged over 20 carrier-envelope phases ($\varphi=2j \pi /20 $). All distributions are normalized to the same maximum count.}
\end{figure}

We compare of our numerical results to experimental ones by inspecting the longitudinal ion momentum distributions. Two groups \cite{Alnaser08,Herrwerth08} reported such measurements for doubly-ionzed neon and argon for different frequencies in the nonsequential regime. The key observation of those reports was that the ''valley'' at zero parallel ion momentum becomes more pronounced with decreasing frequency. Since the lasers used in these experiments were not phase-stabilized, we have to average our momentum distributions over a uniform distribution of carrier-envelope phases to enable qualitative comparison. We picked a peak field amplitude which is in the nonsequential regime for all frequencies ($F_0=0.16$~a.u.) and calculated the distributions for twenty phases $\varphi=2 j \pi /20 $. The resulting averaged longitudinal ion momentum distributions are presented in Fig.~\ref{fig:IonMomDis_averaged}. All distributions are normalized to the same maximum count. In agreement with \cite{Alnaser08,Herrwerth08}, we find that the depth of the valley strongly depends on $\omega$. However, the deepening of the valley with decreasing $\omega$ is not as monotonic as it appeared to be in \cite{Alnaser08}, due to the reduced dimensionality of our model. Nevertheless,the deeper valley at lower frequencies points to a smaller role of RESI in the overall double ionization process.

\section{\label{sec:conclusion}Conclusion}
Despite the continuous development of numerical methods and the growth of computer resources, solving the full–dimensional quantum mechanical problem of atoms interacting with strong short laser pulses is still a
big challenge. Scanning a wide range of laser frequencies (wavelengths) and intensities when calculating ionization yields requires a significant amount of computational resources. Going further, the generation of other experimental observables, e.g. momentum distributions, sets high requirements on the computer memory and disk resources, shifting the focus to the big data analysis. While solutions in this area exist, they will not always be the first method of choice. When the line of compromise between the experiment and the theoretical description is determined at the level of qualitative explanations of the observed phenomenon, simplified models are fully sufficient.

Indeed, this is the case of the problem investigated by us. Using a quantum mechanical
(1+1)-dimensional model, we analyzed the frequency dependence of strong-field double ionization with special regard to experimentally available data, i.e. intensity-dependent single and double ionization yields, ratios of double to single ionization, electron momentum distributions and longitudinal ion momentum distributions. The observed dependence of the single and double ionization yields on the laser frequency is in agreement with available experimental data and semi-classical calculations. In our approach, we used a simplified He-like model, however, accessible data concerns Ar, He, Mg, Ne and Xe atoms. The observed agreement, once again, proves that our model catches the essence of the analyzed phenomena, despite the reduced dimensionality. Usually, the decrease in the ratios of double to single ions with decreasing $\omega$ is attributed to the interplay of wave-packet diffusion, Coulomb focusing and defocusing effects. However, the latter two cannot be accounted for by a reduced-dimensionality model, and wave-packet diffusion is also strongly suppressed there. The observed change in the ratio of double to single ionization is attributed to the smaller role played by RESI at lower frequencies. Next, the influence of the pulse duration on the obtained yields was analyzed, showing that the knee regime is shifted towards lower laser intensities in longer pulses. The obtained results
were corroborated by considering rate equations and exploiting the fact that the ratio of double to single ionization is almost constant in the knee regime. Finally, the study was supplemented with two-electron momentum distributions and longitudinal ion momentum distributions. The transition from the nonsequential to the sequential regime is clearly visible for all studied laser
frequencies. The increase of the momentum space in the momentum distributions with decreasing $\omega$ and increasing $F_0$ is also observed. The smaller role played by RESI at lower frequencies is confirmed by the observed changes in the longitudinal ion momentum distributions with decreasing frequency.

\begin{acknowledgments}
A tribute to the late Bruno Eckhardt with whom we had the honor of working. The above work was initiated and stimulated by him.
\end{acknowledgments}
\bibliography{ref}
\end{document}